\documentclass[12pt,preprint]{aastex}

\tighten
\newcommand\be{\begin{equation}}
\newcommand\ee{\end{equation}}
\newcommand\jcd{Christensen-Dalsgaard}

\usepackage{epsfig}
\begin{document}
\shortauthors{Antia \& Basu}
\shorttitle{Solar rotation during the Cycle 24 minimum}

\title{Solar rotation rate during the Cycle 24 minimum in activity}
\author{H. M. Antia}
\affil{Tata Institute of Fundamental Research,
Homi Bhabha Road, Mumbai 400005, India}
\email{antia@tifr.res.in}
\and
\author{Sarbani Basu}
\affil{Department of Astronomy, Yale University, P. O. Box 208101,
New Haven CT 06520-8101, U.S.A.}
\email{sarbani.basu@yale.edu}

\begin{abstract}
The minimum of solar cycle 24 is significantly different from
most other minima in terms of its duration as well as its abnormally low levels of
activity. Using available helioseismic data that cover epochs from the minimum
of cycle 23 to now, we study the differences in the nature of the 
solar rotation between the minima of cycles 23 and 24.
We find that there are significant differences between the 
rotation rates during the two minima. 
There are differences in the zonal-flow pattern too.
We find that 
the band of fast rotating region close
to the equator bifurcated around 2005 and recombined by
2008. This behavior is different from that during the cycle 23 minimum.
By auto-correlating the zonal-flow pattern with a time shift, we find that
in terms of solar dynamics, solar cycle 23 lasted for a period of 11.7 years,
consistent with the result of Howe et al.~(2009).
The autocorrelation coefficient also confirms that
the zonal-flow pattern penetrates through the convection zone.

\end{abstract}


\section{Introduction}
\label{sec:intro}

Helioseismology has been successfully used to study the dynamics of the solar interior
(e.g., Thompson et al.~1996; Schou et al.~1998, etc.). Both the 
Global Oscillation Network Group (GONG)  and
the Michelson Doppler Imager (MDI) instrument on board {\it SoHO} have observed the
Sun from the beginning of solar cycle 23, and hence, it is now possible to
study how the rotation rate inside the Sun changed with time and solar activity.
In particular, we can now also study whether the solar rotation rate or
zonal flows were significantly different during the minimum of
solar cycle 24 compared to that of cycle 23.

It is well known that the Sun went through an unusual phase of activity
during the last few years with an extended minimum with abnormally low
activity levels that are significantly different from the previous minimum.
This is also reflected in polar magnetic fields which are significantly
weaker during the 2008 minimum as compared to earlier minima
for which detailed magnetogram data are available
(e.g., Wang et al.~2009).
Similarly, {\it Ulysses} measurements find that the high latitude solar wind speed
is slower and density is lower during the latest minimum as compared to the
previous one (McComas et al.~2008). This motivates our study of the differences
that occurred inside the Sun, and seismic data allow us to do that.

There have been a few studies of differences between the
latest solar minimum and the earlier ones for which seismic data exist.
Broomhall et al.~(2009) found significant variations
in frequencies of low-degree modes obtained by
the Birmingham Solar Oscillation Network (BiSON) during the minimum of cycle
24 and that of cycle 23; in particular,
they found that the frequencies were significantly lower during
the cycle 24 minimum as compared to those during the previous one.
Basu et al.~(2010) used the frequency differences between the two minima
to study the corresponding variations in solar structure.
Similar results about frequency differences were found in  data obtained
by GOLF instrument on board {\it SoHO}
(Salabert et al.~2009).
Jain et al.~(2010) have found that the frequencies of 
intermediate degree modes of solar oscillation are also significantly lower during the cycle 24 minimum
as compared to that during the cycle 23 minimum.
Howe et al.~(2009) made the first study of the differences in the
zonal-flow pattern during the two minima. They concluded that the extended
minimum is probably due to the fact that the flow bands were moving more
slowly towards the equator. More data are now available to check the further
progress of these bands and to identify other differences.

Early investigations using subsets of the data that are available now 
had already established that there is a significant temporal variation of the rotation rate of the solar
interior (e.g., Kosovichev \& Schou 1997; Schou 1999; Howe et al.~2000; Antia \& Basu 2000).
The most striking temporal variation shows up in the form of
a system of bands with faster or slower than average rotation
rate in the cuts of rotation profile at a given  depth in the upper layers of the  convection zone.
At low latitudes, the zonal flow bands appear to migrate towards the
equator with time, and appear to be similar in character to the torsional oscillations
discovered in the solar surface rotation rate (Howard \& LaBonte 1980;
Ulrich et al.~1988; Snodgrass 1992).
On the other hand, at high latitudes ($> 45^\circ$) these bands move
poleward with time (Antia \& Basu 2001; Ulrich 2001).
With accumulation of more seismic data it became
clear that these zonal flows  penetrate through most of the convection zone
(Vorontsov et al.~2002; Basu \& Antia 2003, 2006; Howe et al.~2006; Antia et al.~2008).
The zonal-flow pattern is correlated to the magnetic butterfly diagram
(Antia et al.~2008; Sivaraman et al.~2008) and is thought to be intimately connected
with the solar dynamo. It is believed to arise from a nonlinear
interaction between differential rotation and magnetic field (e.g.,
Covas et al.~2000; Rempel 2006).
Hence the properties of solar zonal flows provide a strong constraint on
the solar dynamo models. This makes it crucial to detect
and understand cycle-to-cycle changes of solar zonal flows. This would
provide the necessary input in dynamo models to understand and predict the variation
in activity levels from cycle to cycle.

In this work we study the temporal variations in solar rotation rate with
particular emphasis on the differences between the minima of cycles 23 and 
24.
We also study the changes in the outer shear layer of the Sun.
We also use the available data to estimate the length of solar cycle 23
using a time-shifted auto-correlation.
This  enables us to identify the epoch of latest minimum from the
point of view of solar dynamics and this
may be useful in prediction of activity during the next cycle.

\section{Data and analysis}
\label{sec:data}

We use data obtained by the GONG (Hill et al.~1996) and MDI
(Schou 1999) projects for this work.
These data sets consist of the mean frequency and the splitting coefficients
of different $(n,\ell)$ multiplets.
Only the odd-order splitting coefficients are needed to determine
the rotation rate in the solar interior (e.g., Ritzwoller \& Lavely 1991).
We use 145 data sets from GONG, each set
covering a period of 108 days. The first set starts on
1995, May 7 and the last set ends on 2009, October 31, with a
spacing of 36 days between consecutive data sets.
Thus these sets cover about a year of data leading to the minimum of
cycle 23 as well as several months of data following the minimum of cycle 24.
The MDI data  consist of 69 non-overlapping
sets each obtained from observations taken over a   period of
72 days. The first set begins on 1996, May 1 and the last
set ends on  2010, April 29. The MDI data start close to the minimum
of cycle 23 and do not cover preceding period, while these data cover
a period of about a year following the minimum of cycle 24.

We use the two dimensional Regularized Least
Squares (2D RLS) inversion technique as described by
Antia et al.~(1998) to infer the rotation rate in
the solar interior from each of the available data sets.
The first eight odd-order splitting coefficients, $a_1,a_3,\ldots,a_{15}$, were
used in the inversions. 
For many parts of this investigation we use the  rotational velocity, $v_\phi=
\Omega r\cos\theta$ ($\theta$ being the latitude), rather than the  the rotation rate $\Omega$.

The calculated rotation rate is used mainly to study differences between the minima 
of cycle 23 and that of cycle 24. These  results also allow us to study
the general nature of the time variation of the solar rotation rate.
In particular, we concentrate on the changing nature of the solar
zonal-flow pattern and the outer shear layer of the Sun.
It is well known that 
there is a strong shear layer near the solar surface 
where rotation rate generally increases with depth. 
 The radial gradient in the outer shear layer is, however, not as
strong as that in the tachocline.
Antia et al.~(2008) studied the time-variations in
the radial gradient of
rotation rate in the solar interior, but the results in the outer layers
were not particularly reliable. In this work we try to study gross
features of the shear layer, i.e., the extent of shear layer and the net
change in the rotation rate in this layer.
This also allows us to look for changes between the two epochs of interest. 
We define 
shear layer at any latitude as the layer where the radial gradient exceeds 20\% of its
value near the surface. This definition was adopted because the location
of the layer where the radial gradient vanishes is somewhat uncertain
because of the flatness of the rotation profile near the maximum in the rotation rate.
Furthermore, at high latitudes the rotation rate sometimes keeps increasing throughout
the convection zone, and hence, our definition allows the shear layer to be defined as the region
near the surface where the gradient is significant.
In addition to the  depth, we also find the increase in rotation rate, $\Delta\Omega$,
between $r=R_\odot$ and $r_s$, the base of the shear layer as defined above.

To study solar zonal flows, we use the inverted rotation rates for each epoch
and  take the time average of the results over a period that covers a solar cycle at each latitude
and depth.  We treat the GONG and MDI sets separately and hence we  obtain
one average result for GONG sets and another for  MDI sets.
To obtain the time-varying component of rotation, we subtract the  mean from the
rotation rate at any given
epoch. Thus,
\begin{equation}
\delta\Omega(r,\theta,t)=\Omega(r,\theta,t)-
\langle \Omega(r,\theta,t)\rangle,
\label{eq:zonal}
\end{equation}
where $\Omega(r,\theta,t)$ is the rotation rate as a function of
radial distance $r$, latitude $\theta$ and time $t$.
The angular
brackets denote the average over a solar cycle.
Since the available data cover a period longer than the solar cycle, the
temporal average is not taken over the entire length of data since that could
skew the average towards the value of the rotation rate during the solar minimum. Instead, we 
use an estimate of the length of the solar cycle, $T_0$,
and beginning with the first set, average over all data sets over this interval.
The estimate of the solar-cycle length is obtained using  an autocorrelation technique as explained
below. As mentioned earlier we generally show the result for zonal flow
velocity $\delta v_\phi=\delta\Omega r\cos\theta$.
The time-averaged rotation rate, as well as the residual rotation rate, is 
calculated separately for the GONG and MDI data sets. By this process systematic
differences between GONG and MDI data sets (see e.g., Schou et al.~2002) are
largely canceled out and there is a good agreement between the
zonal-flow velocities, $\delta v_\phi$, obtained using the two data sets.

Since the data cover a period of more than one  solar cycle,  and since  we expect
the zonal-flow pattern to vary with the solar-cycle period,
we try to estimate a `dynamical' length of cycle 23 using the zonal
flow results.
For this purpose we calculate
the correlation of the zonal-flow pattern with itself, but  with a time
delay:
\begin{equation}
C(T,r)=\frac{\sum_{i,j}\delta v_\phi(r,\theta_i,t_j)\delta
v_\phi(r,\theta_i,t_j+T)}{\sqrt{(\sum_{i,j} \delta v_\phi(r,\theta_i,t_j)^2)
(\sum_{i,j} \delta v_\phi(r,\theta_i,t_j+T)^2)}}\;.
\label{eq:cor}
\end{equation}
The summation is taken over all latitudes that we have inversion
results for and over all admissible epochs. This definition is similar to that
used by Howe et al.~(2009), except that we have summed over time also,
maintaining constant time difference. On the other hand, unlike Howe
et al.~(2009) we have not summed over radial points. Considering the
narrow range of $r$ values used by them, this summation will not make
much difference and our value at $r=0.98R_\odot$ would represent the
average over the outer layers that they use.
The correlation coefficient was calculated separately for the GONG and MDI results.
The rotation-rate inversion results were calculated at  latitudes from
$0^\circ$ to $88^\circ$ in steps of $2^\circ$. Furthermore, since we use all the available data sets for a given project,
time $t_j$ is the epoch of a given set.
As mentioned above, we calculate the correlation coefficient separately for
different depths. To avoid interpolation we only use values of $T$ that are
integral multiple of spacing between consecutive data sets. The number of terms
in the summation depends on $T$ --- there are fewer
pairs of time $(t_j,t_j+T)$ available for larger values of $T$. The maximum permissible value of $T$
is the length of data set, which is about 14 years.
At each depth, the maximum of the correlation function can be
expected to occur when $T$ is equal to the solar cycle length, $T_0$, and hence we 
look for the value of $T$ for which $C(T,r)$ is the largest. This is
justified since the maximum correlation is found
to be close to unity in the near-surface layers. Since the behavior of the  zonal-flow pattern in the
low-latitude regions is different from that in the high latitude regions,
we also  calculate the correlation
coefficient separately for the two regions. We define the boundary of the two regions to be
at the latitude of $ 45^\circ$. The exact location
of the defining boundary does not affect the result as long as the low-latitude
region encompasses the equator-ward flow and the high-latitude region contains most
of the pole-ward flows. 
It is obvious that we need to iterate on the process, since the value of $T_0$ determines the
number of sets we average over time to determine  $\langle v_\phi\rangle$ that is used
to estimate $T_0$. The process converges rapidly since the temporal average is
not very sensitive to $T_0$.

We use the estimated period, $T_0$ to fit the temporal variation with a periodic
function of the form
\be
\delta v_\phi(r,\theta,t)=\sum_{k=1}^{k_{\rm max}} a_k(r,\theta)\sin(k\omega_0 t)
+b_k(r,\theta)\cos(k\omega_0 t)=\sum_{k=1}^{k_{\rm max}} A_k\sin(k\omega_0 t+
\phi_k)\;.
\label{eq:harm}
\ee
In the above equation, $\omega_0=2\pi/T_0$ is the basic frequency of the fundamental component. 
We use two harmonics for these fits (i.e., $k_{\rm max}=3$).
Similar fits were used by Vorontsov et al.~(2002), Basu \& Antia (2003) and others.
The fits are calculated for each radius and latitude. The results for amplitudes, $A_k$
and phase $\phi_k$  were shown by
Antia et al.~(2008) for a somewhat smaller data set and we will not repeat them here. However, we do
show the fits for
a few cases to demonstrate that the zonal-flow pattern is indeed 
approximately periodic.

\section{Results and discussion}
\label{sec:results}

\begin{figure}
\plotone{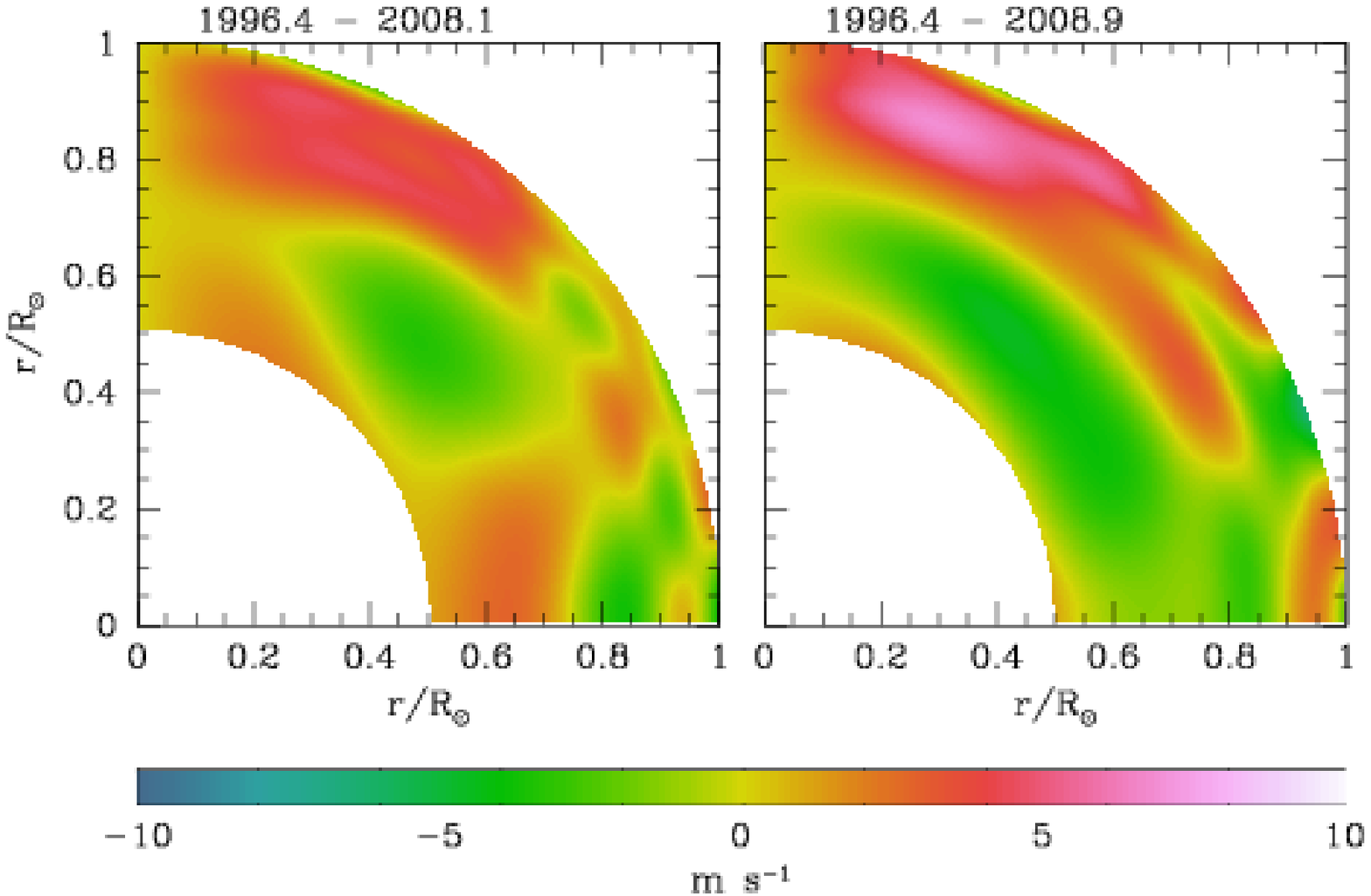}
\caption{The difference in rotation velocity between the minimum of
cycle 23 and that of  cycle 24. We show the difference between the results of
1996.4 and 2008.1 (the cycle 24 'dynamical' minimum) on the left and
1996.4 and 2008.9 (the cycle 24 minimum as per magnetic indices) on the right.
The vertical axis represents the rotation axis.}
\label{fig:rotdif}
\end{figure}

We find that the solar rotation rate during the minimum of cycle 24 was
significantly different from that of cycle 23, regardless of how the 
cycle 24 minimum is defined. We show the differences in
rotation velocity $v_\phi$ between the two minima  obtained using GONG
data in Figure~\ref{fig:rotdif}.
We assume the cycle 23 minimum to be at 1996.4 and
we show differences for two different estimates of the cycle 24 minimum:
one dynamical (2008.1) and one from conventional activity indices (2008.9).
To improve statistics, we have averaged the results for 5 data sets
around each epoch before subtracting.
One can see that there are differences in all parts of the solar interior
for which we have reliable inversion (technically, we cannot invert near the poles,
however the $\cos\theta$ factor in $v_\phi$ make the difference tend to zero
at the poles).

\begin{figure}
\epsscale{.80}
\plotone{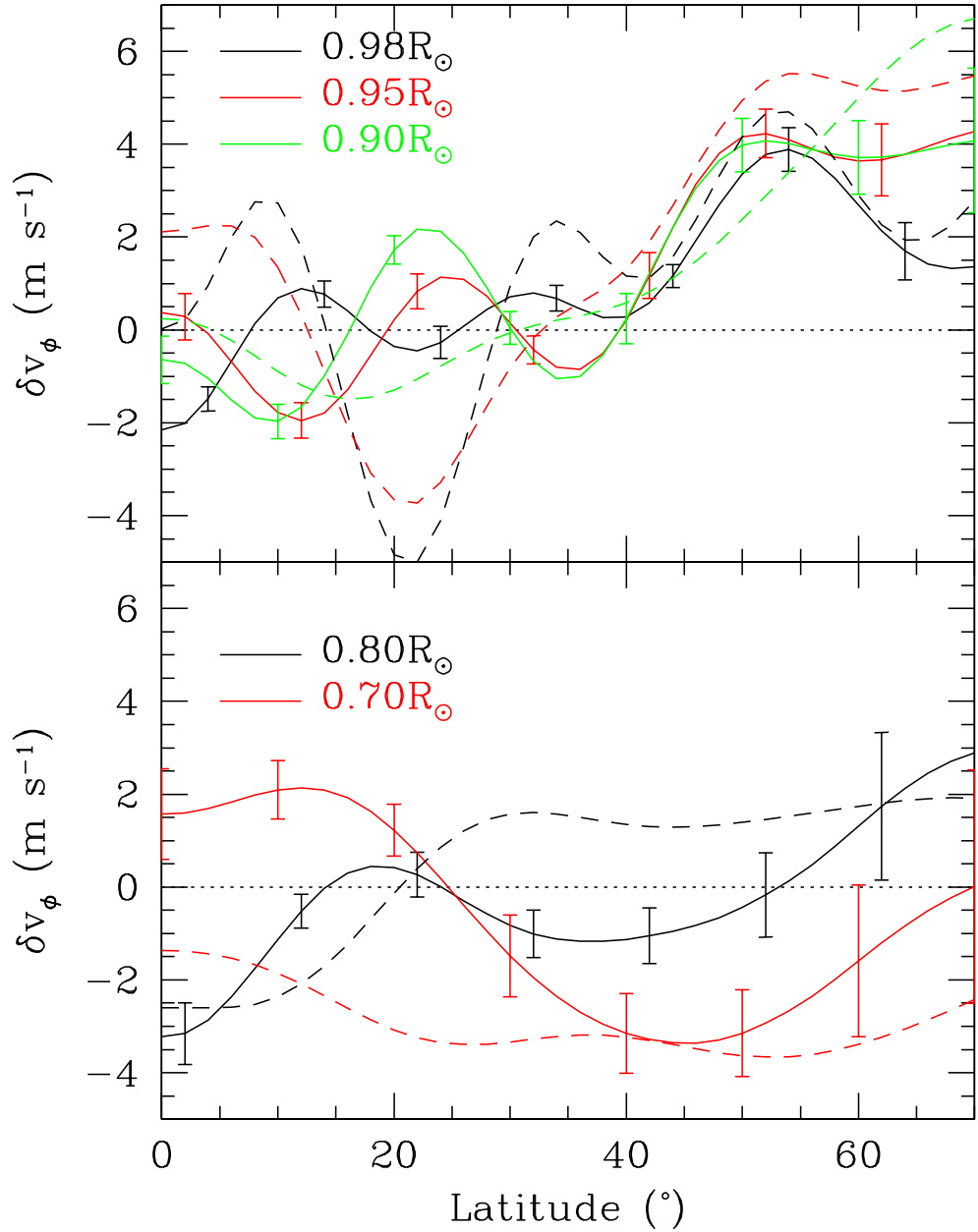}
\caption{
The same as Fig.~\ref{fig:rotdif}, but showing cuts at constant radius
 as a function of latitude. The continuous
lines show the difference between 1996.4 and 2008.1, while
the dashed lines show the same between 1996.4 and 2008.9.}
\label{fig:difcut}
\end{figure}

In order to gauge the significance of the velocity differences, we show radial cuts of the
results in Figure~\ref{fig:difcut} so that we can plot the results along with 
error-estimates. We show the results for different parts of the solar
convection zone, paying particular attention to the two dynamically significant parts of the
Sun --- near the tachocline (believed to be the seat of the solar dynamo) and
the outer shear layer (a region which could support local turbulent dynamos).
As can be seen, there are substantial differences
between the rotational velocities at the two minima.
There are differences
in the tachocline region too. Since the tachocline is where the solar dynamo is assumed to
exist, this result may have implications for solar dynamo models.
Note however, that the differences in the tachocline region are of the order of $3\sigma$
or less and hence their significance is not completely clear.
Additionally, there are changes in the outer shear layer too. 
These differences have a somewhat higher significance level.
The difference
in the outer shear layers probably reflect the differences in the zonal flow
velocities that we describe later.

\subsection{The outer shear layer}

\begin{figure} 
\epsscale{0.9}
\plotone{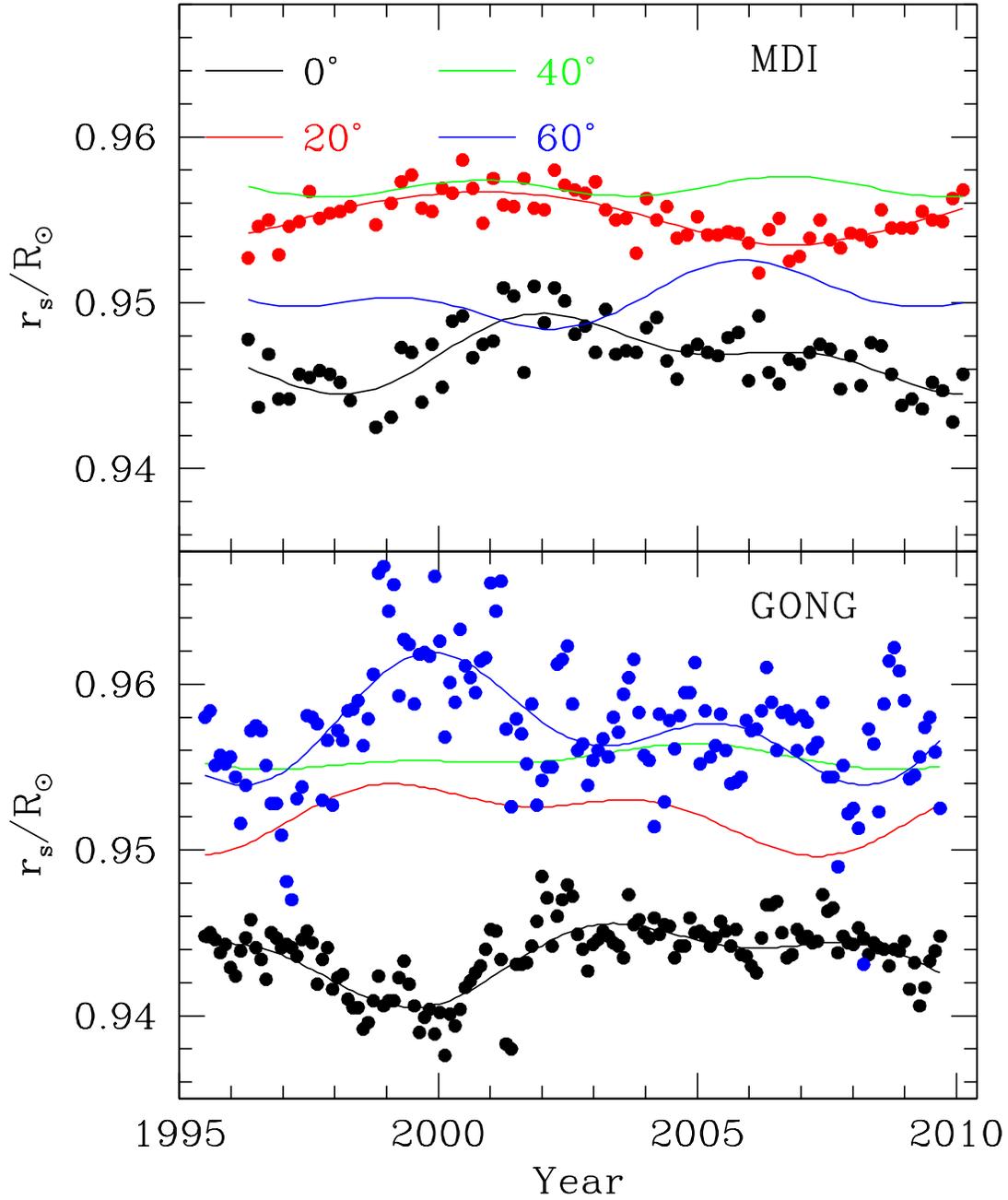}
\caption{The radial distance, $r_s$ of the base of the outer shear layer
as a function of time at a few selected latitudes as obtained using GONG
and MDI data.
The points show the result for individual data sets and the lines mark the
fits for a periodic signal. Points are shown only for two
latitudes for the sake of clarity.}
\label{fig:shear}
\end{figure}

\begin{figure} 
\epsscale{0.9}
\plotone{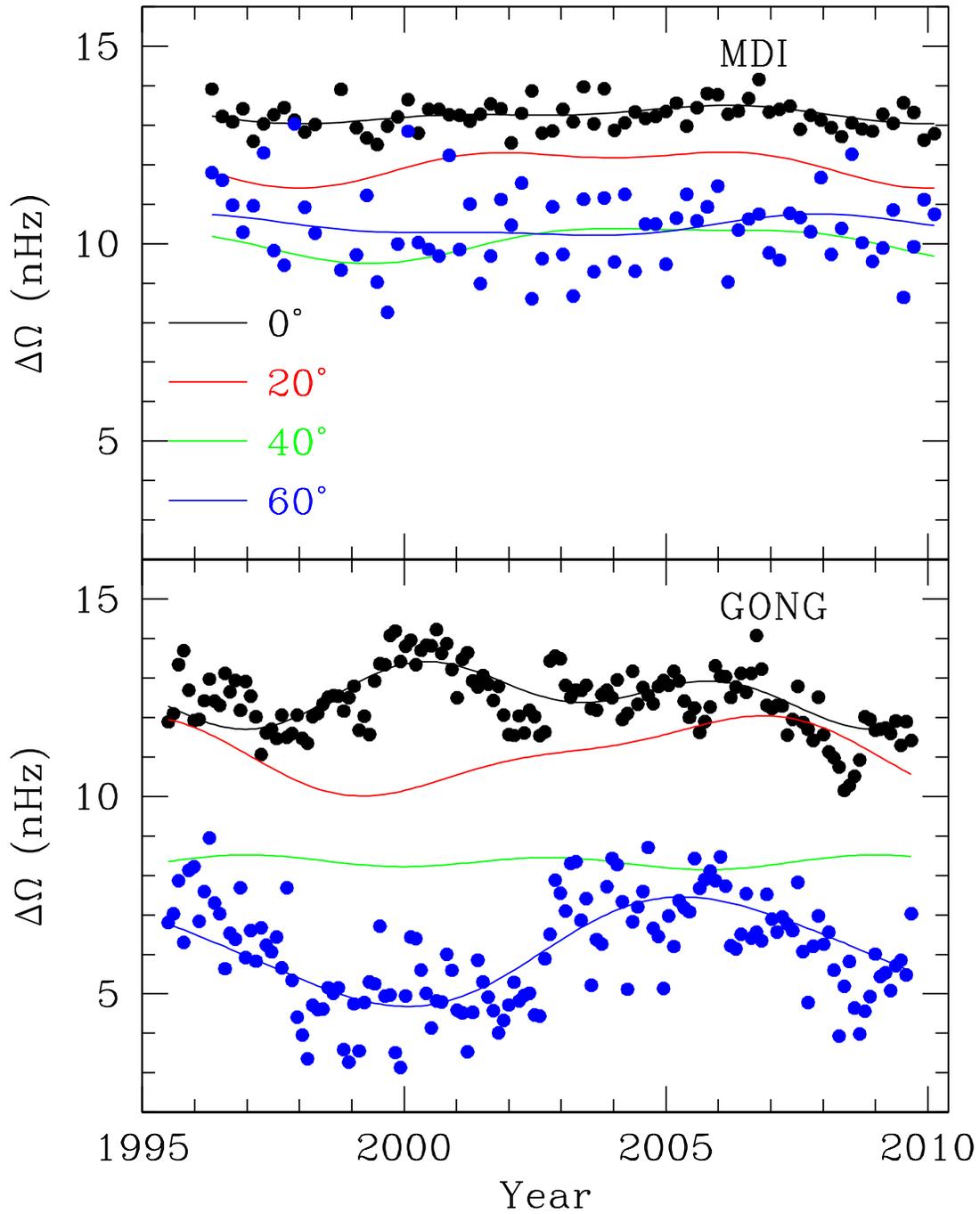}
\caption{The increase in rotation rate, $\Delta\Omega$, across the outer shear layer are shown
as a function of time at a few selected latitudes as obtained using GONG
and MDI data.
The points show the result for individual data sets and the lines mark the
fits for a periodic signal. For clarity,  points are shown only for two
latitudes.}
\label{fig:shearom}
\end{figure}

Given that we find differences in the velocities in the outer shear layer
between the two minima, we investigated whether or not the change in the rotation
rate across the
shear layer and the extent of the shear layer change with time, and
whether these quantities are different for the two minima.
Using the definition of the shear layer laid down in \S~\ref{sec:data},
 we calculate the radial distance, $r_s$, to the base of
the shear layer and the increase in rotation rate, $\Delta\Omega$, between
$r=R_\odot$ and $r=r_s$ for each data set.  The $r_s$ values for all datasets are shown in
Fig.~\ref{fig:shear}, while Fig.~\ref{fig:shearom} shows the results for
$\Delta\Omega$. In both figures GONG results show some temporal variations, but the
MDI results do not.

To get a better idea of whether or not there are  temporal variations in the
properties of the outer shear layer,
we subtract the temporal mean from the values for each data set to get the
residuals of $r_s$ and $\Delta\Omega$ in the manner of the zonal flow analysis.
The residuals are then fitted to a periodic function similar to that of
Eq.~\ref{eq:harm}. 
The lines in Figures \ref{fig:shear}, \ref{fig:shearom}
are these fits for different latitudes.

\begin{figure} 
\epsscale{1.0}
\plotone{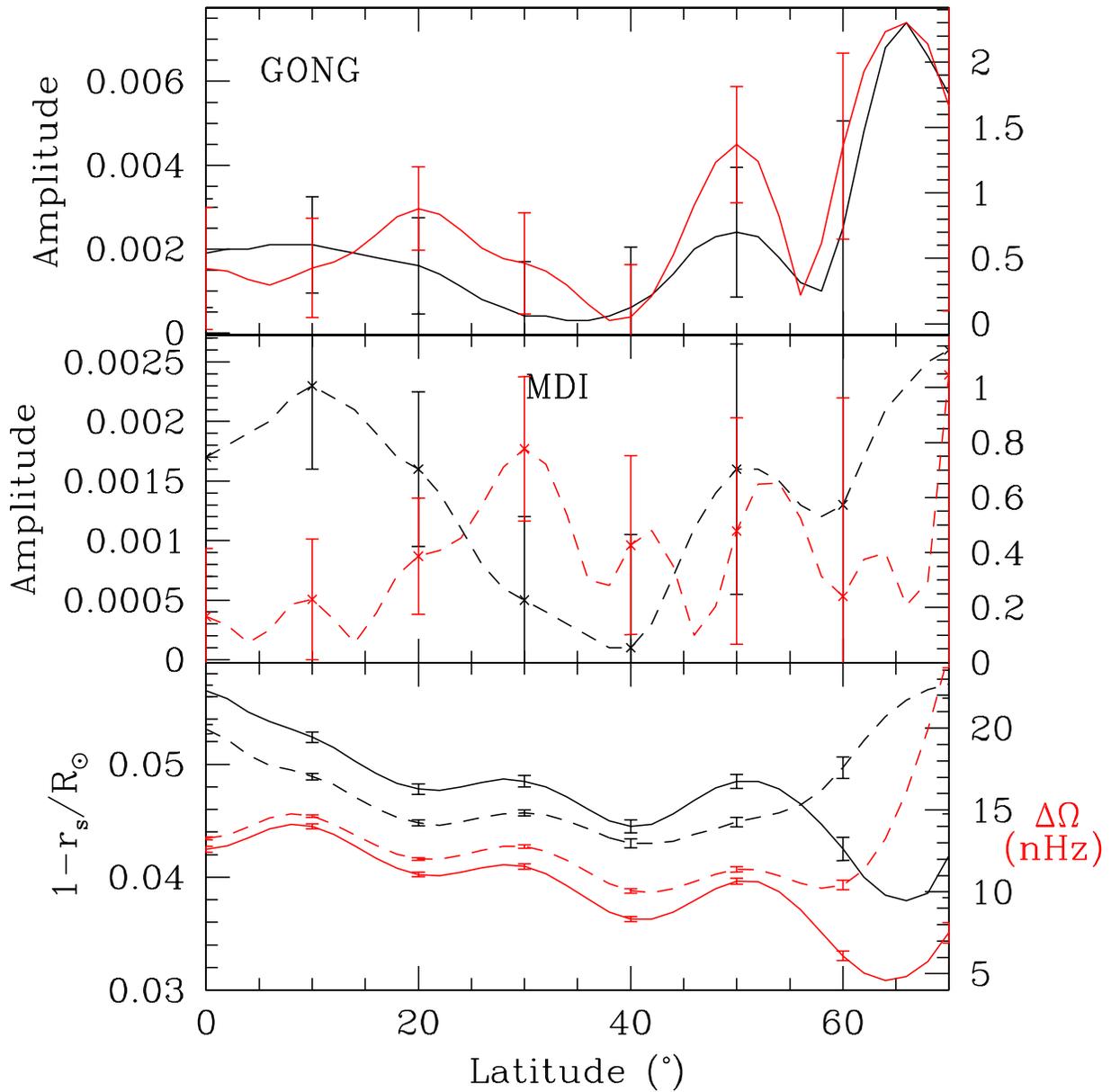}
\caption{The upper two panels show the
 amplitude of the  time-varying component
of the radial distance $r_s$ of the base of the outer shear layer (black) and 
the increase in rotation $\Delta\Omega$ across the layer (red) 
for GONG (topmost panel) and MDI data (middle panel).
The scale for $r_s$ is on the left and that for $\Delta\Omega$ on the right.
The lowest panel shows the time-averaged value of  $r_s$ and
$\Delta\Omega$.
GONG results are shown as continuous lines while MDI results as dashed lines.}
\label{fig:shearlat}
\end{figure}

The amplitude of the time-varying component for the fundamental period of $T_0$,
as well as the time averaged value of
$r_s$ and $\Delta\Omega$
are shown as a function of latitude
in Figures~\ref{fig:shearlat}.
There are some differences even in the
mean values for the GONG and MDI data. This is due to known differences between
the two data sets (Schou et al.~2002). 
For both GONG and MDI it can be seen that the amplitude
of the time-varying component is small and comparable to the error estimates at
each latitude. Hence it is not clear if there is
any significant temporal variation in the extent of the outer-shear layer
or for that matter in the change of rotation rate across the layer.
Thus it appears that although flow velocities at a given radius in the
shear layer differ for the two minima, the change in velocity across the
shear layer was nearly the same for both minima.

Antia et al.~(2008) had studied temporal variations of the radial gradient
of the solar rotation rate, but did not find any clear variation in the near
surface layers. This is consistent with the fact that we do not find any
significant temporal variations in the gross properties of the shear layer.
This may be expected from the fact that the zonal-flow pattern shown in
the upper panel of Figure~\ref{fig:zonal} does not change much with depth
in the shear layer and hence the variation in radial gradient is much
smaller. Below the shear layer the phase of zonal-flow pattern changes
giving some temporal variation in the radial gradient.

\subsection{Zonal flows}

\begin{figure} 
\epsscale{1.0}
\plottwo{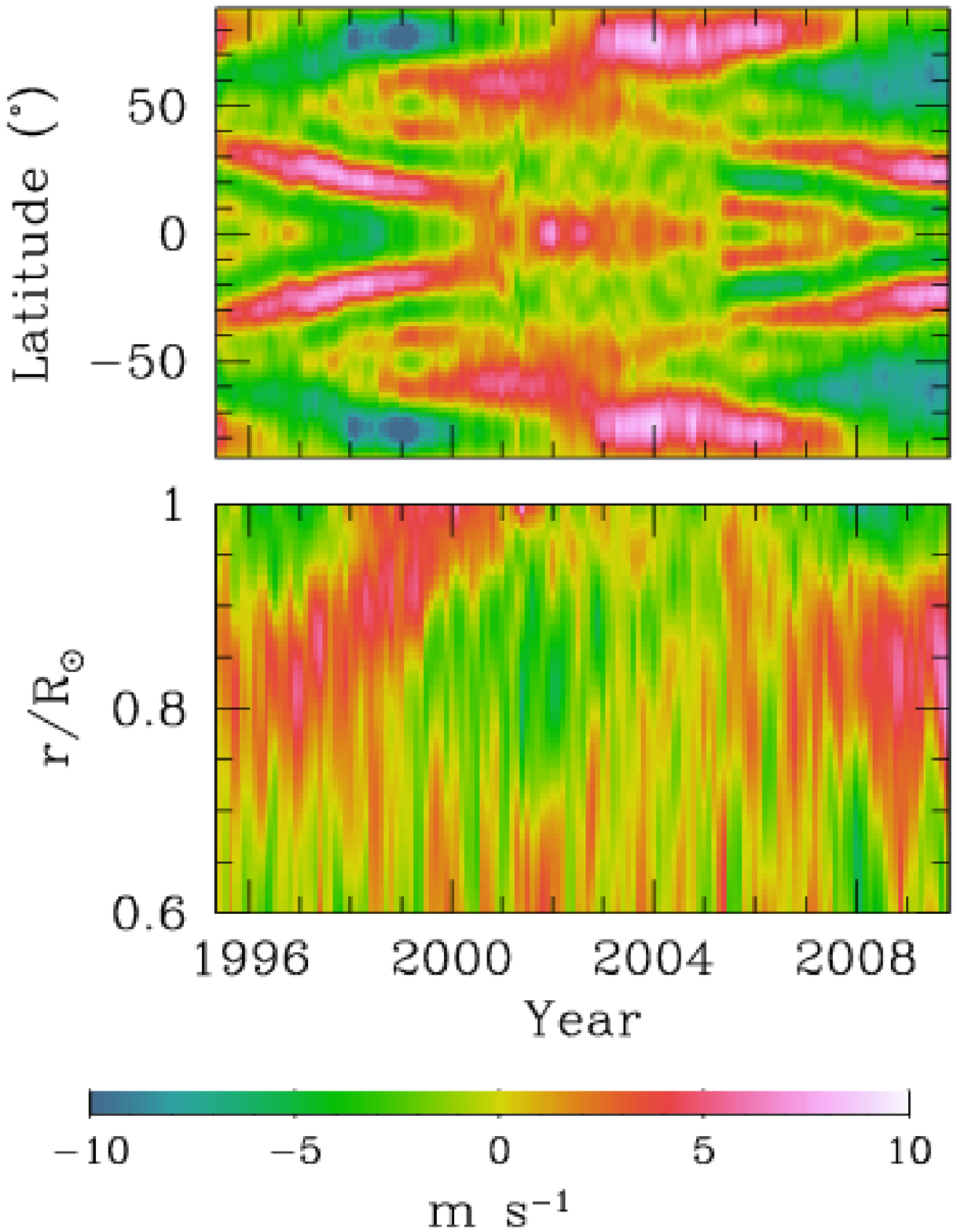}{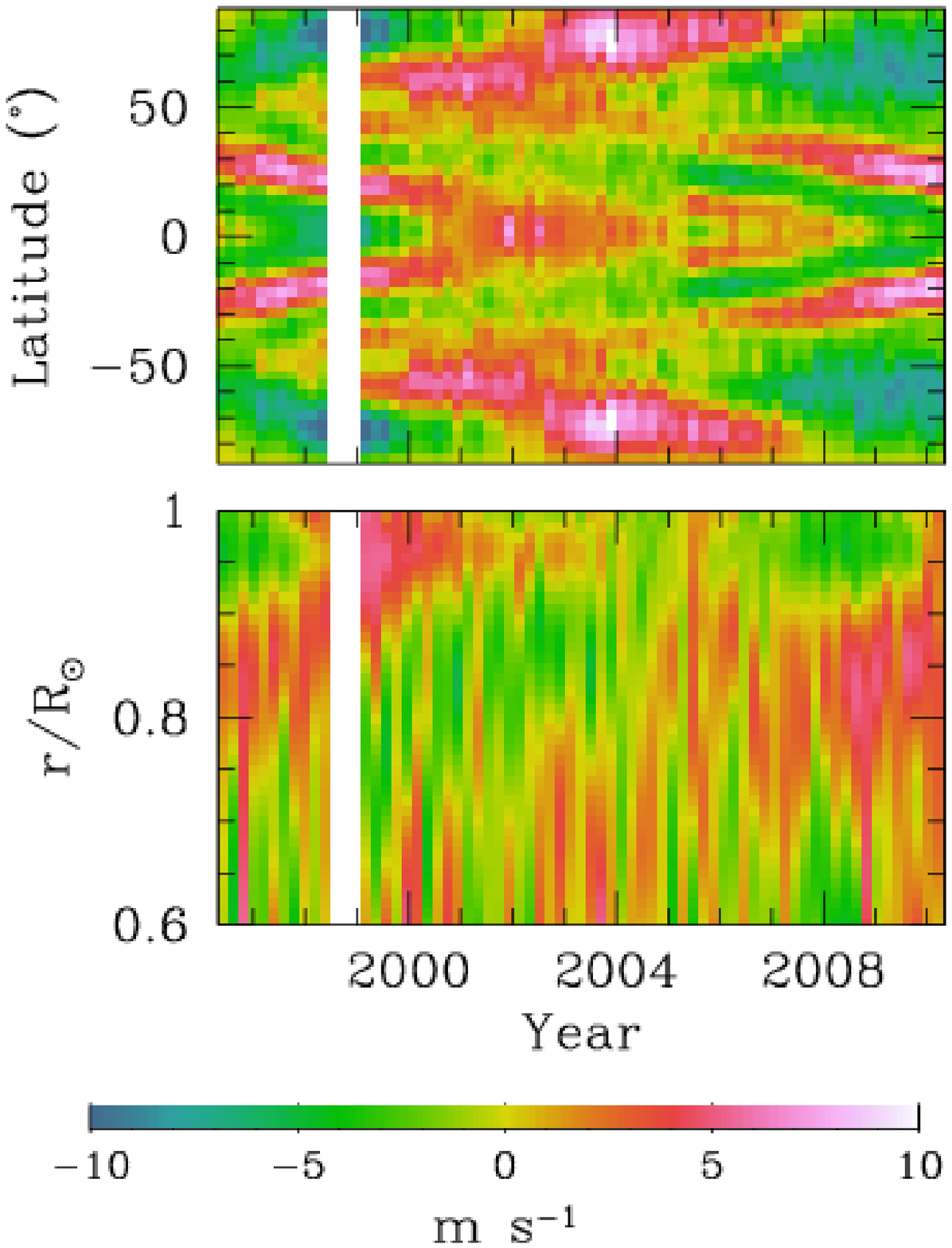}
\caption{ Rotational-velocity residuals, $\delta v_\phi$,
obtained using GONG (left panels) and MDI (right panels) data.
The top panels show the latitude v/s time plot at $r=0.98R_\odot$,
while the lower panels show the depth vs time plot for $15^\circ$ latitude.}
\label{fig:zonal}
\end{figure}

To study the temporal variation in the rotation velocity
we calculated the residual rotation velocity $\delta v_\phi$ separately for
GONG and MDI data sets. A sample of some results obtained using
the GONG and MDI data are shown in Figure~\ref{fig:zonal} which shows cuts at
$r=0.98R_\odot$ and latitude of $15^\circ$. The expected, and well-known, pattern of
zonal flows with bands of faster and slower than average rotation
moving towards the equator at low latitudes is clearly seen. This is similar
to the torsional-oscillation pattern seen both in Doppler measurements at
the solar surface
(Howard \& LaBonte 1980; Ulrich 2001) as well as in other helioseismic studies
(e.g., Schou 1999; Vorontsov et al.~2002; Basu \& Antia 2003, 2006; Howe et al.~2006, Antia et al.~2008).
The high-latitude, poleward flow, pointed out 
by Antia \& Basu (2001) using only a subset of the data, is also seen.

In this study we are more interested in the differences in the zonal-flow
patterns between the last two minima, rather than the gross properties of the flow and hence we
take a close look at the evolution of the pattern. Inspection of Figure~\ref{fig:zonal}
shows that the lowest latitude band of faster than average rotation
merged to form one band near the equator around 2000, but split
again around 2005 to merge-again around 2007. This band slowed down and disappeared around the
beginning of 2009 and the zonal flow at the equator became retrograde.
Both GONG and MDI data show this behavior.
Looking at the bands near the beginning of cycle 23 in GONG data, it
appears that the split bands merged at the beginning of 1996 and the band 
disappeared soon after that just as the activity minimum was reached. Furthermore, this part of
the band is not particularly well defined during the minimum of cycle 23. 
During the recent extended minimum the migration of the intermediate-latitude
band of faster rotation towards the minimum was slower.
This has been pointed out by Howe et al.~(2009). However, it should be noted that
they did not detect the splitting and re-merging of the low-latitude
band. Nevertheless, their results also show some non-smooth behavior
in the equatorial band.
Although we do not have much data before the minimum of cycle 23, it appears
that during cycle 24, the equatorial band of faster than average rotation has extended for
longer period after the two bands merged. The end of this band around 2009.0
is close to the estimated minimum from conventional activity indices.
As far as the high-latitude poleward flows are concerned,
the band of faster-than-average rotation ended around 1996.5 close to the
period of minimum, while the next band ended by 2008.0, somewhat before the
minimum as defined by usual solar activity indices.
Thus while the end of low latitude band coincides with the time of
minimum as inferred from conventional activity indices,
the end of high latitude fast band around 2008.0 matches
with the estimate of minimum as obtained from zonal flows (Section \ref{sec:cycle}).
Considering that equatorward movement of mid latitude band is slower during
cycle 24 and
given the correlation between these bands and the
emergence of active regions (see Antia et al.~2008), it is not surprising that 
cycle 24 was delayed.
From Figure~\ref{fig:zonal} it also appears that the intermediate-latitude 
prograde band at the beginning of  cycle 23 was at a slightly higher latitudes 
than the corresponding band at the cycle 24 minimum. The correlation between the
bands and sunspot positions would lead one to expect that the first sunspots
for cycle 23 would have appeared at a higher latitude than those for cycle 24. This
does indeed seem  to be the case.

\begin{figure}
\epsscale{.70}
\plotone{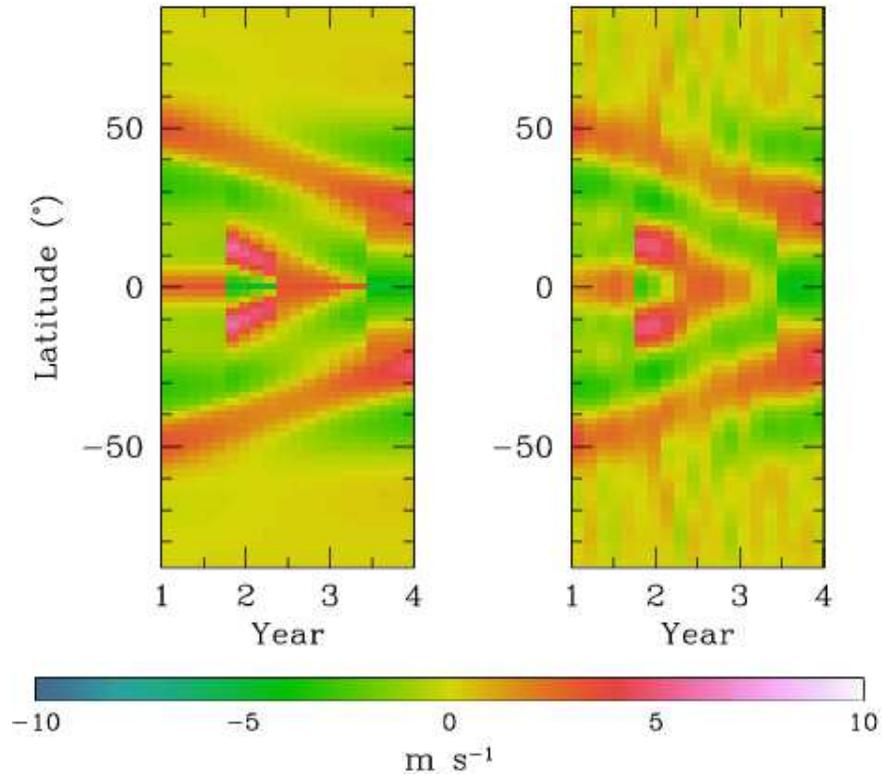}
\caption{The zonal-flow velocity as a function of time at different 
latitudes at $r=0.98R_\odot$ for an artificial data set. The left
panel shows the input velocity, while the right panel shows the results
obtained by inversions when errors consistent with those in a typical MDI data set
were added to the calculated splitting coefficients.}
\label{fig:art}
\end{figure}

In order to verify that the splitting and remerging of equatorial band 
is not an artifact of the lack of latitudinal resolution of the
inversions, we have done some experiments with artificial data
to verify whether or not the latitudinal resolution of our inversion is
adequate.  For this purpose we generated
a sequence of rotation profiles at intervals of 0.2 years, each with 
bands of faster and slower than average rotation superposed on the
average rotation rate. The rotational splittings were calculated for each
of these rotation profiles and random errors consistent with those in an
MDI data set were added to these splitting coefficients and inversions
were performed using the mode set in an MDI data. The
regularization parameter in the 2D RLS technique was the same as that used
for the real observations. The results of this exercise are shown in Fig.~\ref{fig:art},
which compares the input zonal flow profile that with that obtained
by inversions. The bands were generated with a maximum of $\delta\Omega=1$
nHz and the equatorial bands had a half-width of $5^\circ$.
These values are comparable to those seen in the observed split bands.
It is clear
that we are able to infer these bands reliably through our inversions.
We have done this exercise with bands of different widths and
amplitudes of $\delta\Omega$ to find that even for an
amplitude of 0.5 nHz, it is possible to recover the
split bands. However, if the half-width is reduced to $3^\circ$, the inversions
give only a hint of splitting in these bands.
Thus the splitting seen in Fig.~\ref{fig:zonal}
is most likely to be real. This conclusion is further strengthened by the fact
that independent data sets from GONG and MDI show the splitting of the
equatorial band at the same time.

In addition to differences in the latitudinal pattern, we find differences
in the radial pattern of the flow too, though the differences are more subtle.
Antia et al.~(2008) had shown that at low latitudes the 
alternating bands of fast and slow rotation rise upwards through the convection zone
with time as is shown  in the lower panel
of Figure~\ref{fig:zonal}. The panel shows a cut at $15^\circ$ latitude.
The first high-speed band reached the surface around 1998.
The next band of fast rotation appears
to be more well defined than the first one. 
The band appears to reach well into the tachocline 
at its deepest point, this was not the case during the cycle 23 minimum.
This band appears to have started its upward motion in 2009.
It should reach the surface in the next few years.

\begin{figure}
\epsscale{.80}
\plotone{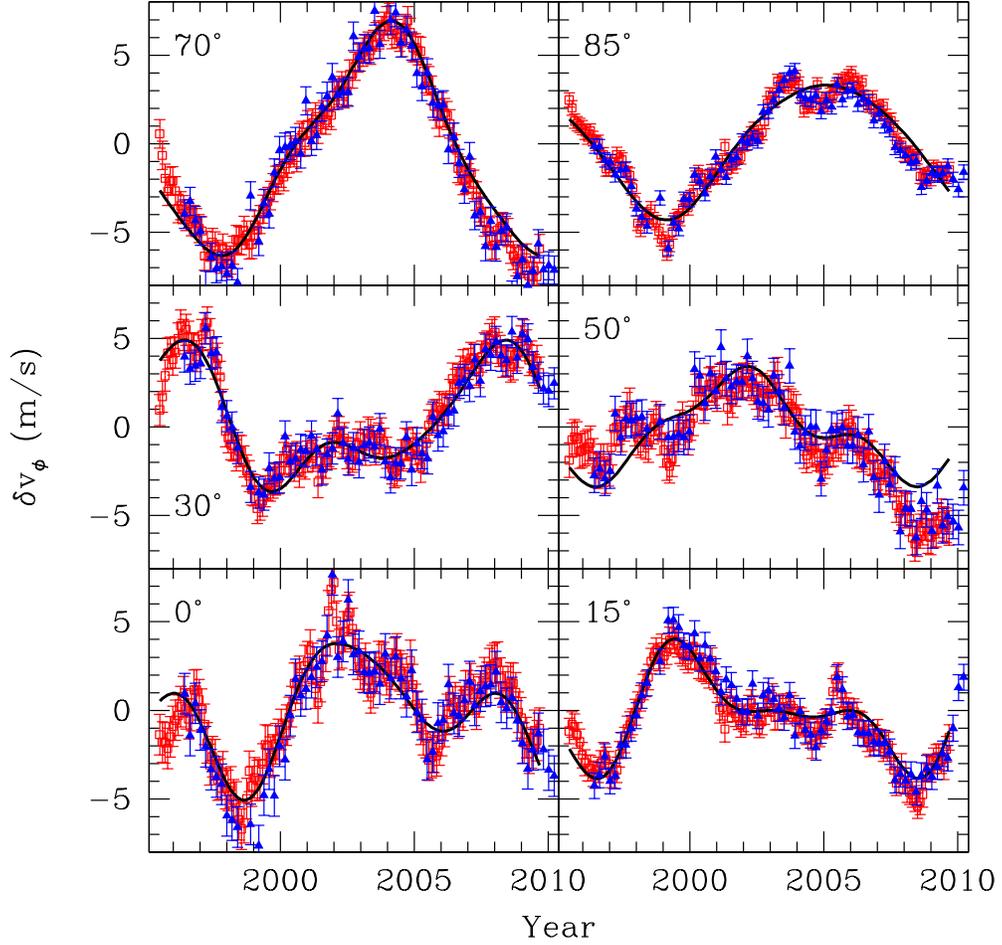}
\caption{The zonal-flow velocity as a function of time at different 
latitudes at $r=0.98R_\odot$. The latitudes are marked in each panel.
The red points show results obtained with GONG data, while blue points
show results with MDI data. The black lines show the fits
using  Eq.~(\ref{eq:harm}) with $k_{\rm max}=3$ to GONG data.}
\label{fig:latcol}
\end{figure}

\begin{figure}
\epsscale{.80}
\plotone{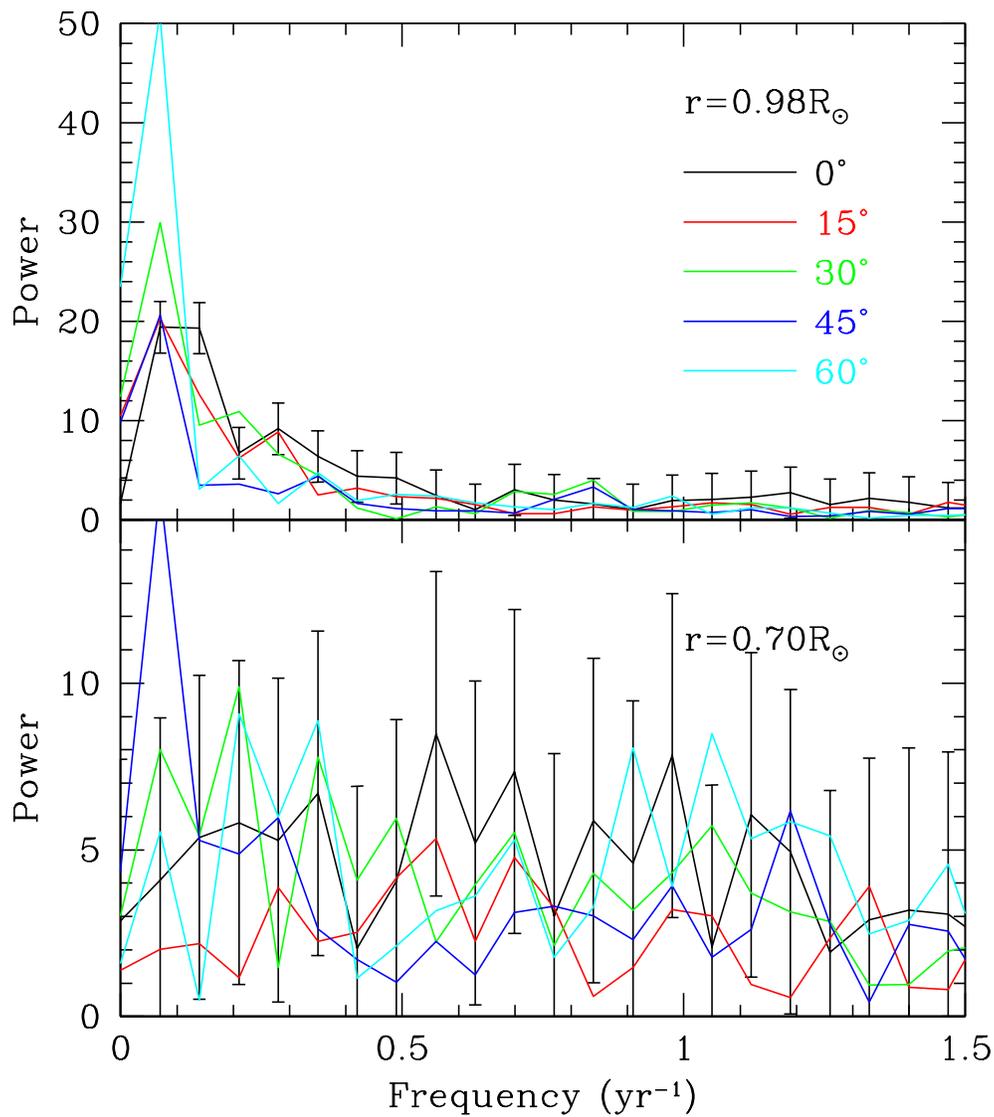}
\caption{The power spectra obtained by Discrete Fourier Transform of
$\delta v_\phi$ at different
depths and latitudes as marked in the two panels obtained using GONG data.
For clarity errorbars are shown only for one latitude.}
\label{fig:dft}
\end{figure}

Figure~\ref{fig:latcol} shows the zonal flow velocity as a function of time
at $r=0.98R_\odot$ for a few selected latitudes.
It can be seen that there is good agreement between the GONG and MDI results.
The fits using Eq.~(\ref{eq:harm})
are also shown and it is clear that the pattern is roughly periodic.
To estimate if any periodicities other than the solar cycle are present,
we take a discrete Fourier Transform of the time series at each latitude
and radius. The results for $r=0.98R_\odot$ and $r=0.7R_\odot$ at a few
selected latitudes are shown
in Figure~\ref{fig:dft}. The higher peak corresponds to the period of the solar cycle and
shows up in the near surface regions. There is no other significant peak
in the power spectra at any depth. There are no significant peaks in the tachocline region ($r=0.70R_\odot$).
Very few latitudes even have a peak corresponding to the solar cycle. Thus
there does not appear to be any significant periodic  temporal variation of the tachocline
rotation rate.

\subsection{Estimate of solar cycle period}
\label{sec:cycle}

\begin{figure} 
\plotone{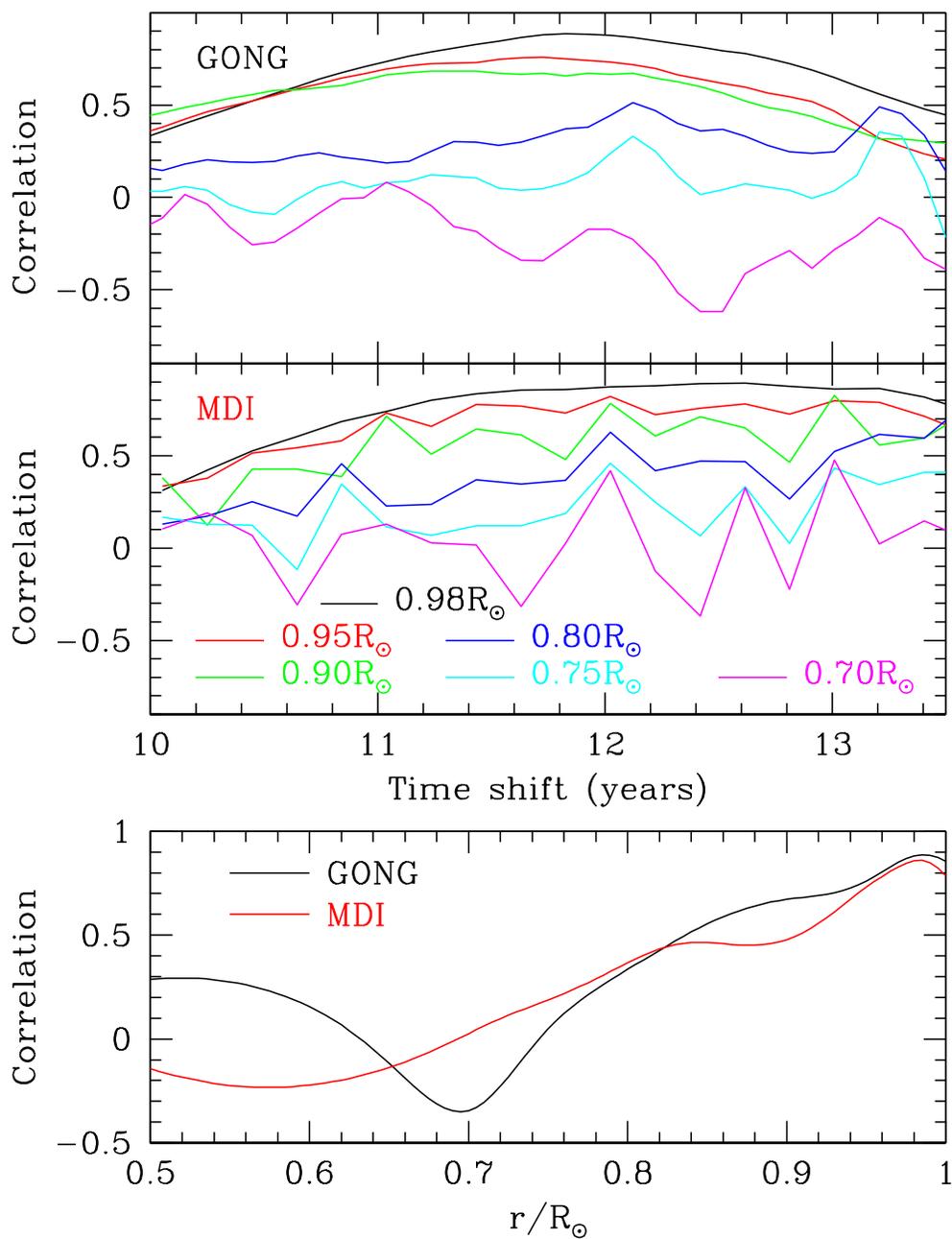}
\caption{Autocorrelation coefficient for GONG and MDI at different depths
is shown as a function of time-shift. The top panel shows the results for
GONG while the middle panel shows those from MDI. The lowest panel shows
the autocorrelation for a time-shift of 11.8 yrs as a function of depth.}
\label{fig:cor}
\end{figure}

Given that both Figures~\ref{fig:latcol} and \ref{fig:dft} appear to show that the
zonal flows are cyclic, we investigate this further. 
To estimate the period, i.e., the length of solar cycle 23, we calculate the
correlation coefficient as defined by Eq.~(\ref{eq:cor}) and the results
are shown in Figure~\ref{fig:cor}. It can be seen that in near-surface layers
the maximum value of $C(r,T)$ is larger than 0.9, giving a very tight correlation.
However, the peak is rather broad and it may be difficult to define the
exact length of solar cycle. 
For GONG data the peak in the correlation-coefficient
is around $T=11.8$ years near the surface and slowly shifts towards lower
values with increasing  depth. Below about $r=0.9R_\odot$ the peak shifts back towards
higher values of $T$ while the peak of the correlation-coefficient decreases with 
increasing depth.
A similar estimate of the period is obtained if instead of using the correlation coefficient
we use the minimum in mean square differences in the rotation
rate between two epochs separated by $T$.
Near the convection zone base there are multiple peaks in  the plot of correlation coefficient
as a function of $T$, the 
 maximum value of $C(r,T)$ is rather low as well. Thus it is not clear if any period can
be defined at these depths or whether there is any significant variation in
the rotation rate. 
For MDI data the behavior is similar, except that the curves are not very
smooth. 
There are some differences
between GONG and MDI data in autocorrelation function, but as will be seen
later these are due to differences in the high-latitude regions.

The lowest panel in the figure shows the correlation
coefficient as a function of depth for $T=11.8$ years. It can be seen
that from a peak value of around 0.9 near the surface, the correlation
coefficient decreases slowly with increasing depth till close to the base
of the convection zone. Near the base of the convection
zone there is a sharp dip and $C(r,T)$ becomes negative for flows obtained with GONG data.
The position of this
dip may be interpreted as a lower limit on the depth 
to which the zonal-flow pattern penetrates into the solar interior. This gives another indication
that the zonal flows penetrate to depths near the base of the convection zone
as has been
claimed by Vorontsov et al.~(2002), Basu \& Antia (2003), Antia et al.~(2008) and others.
For MDI data also the results are similar, though the dip near the base
of the convection zone is less marked.

\begin{figure} 
\plotone{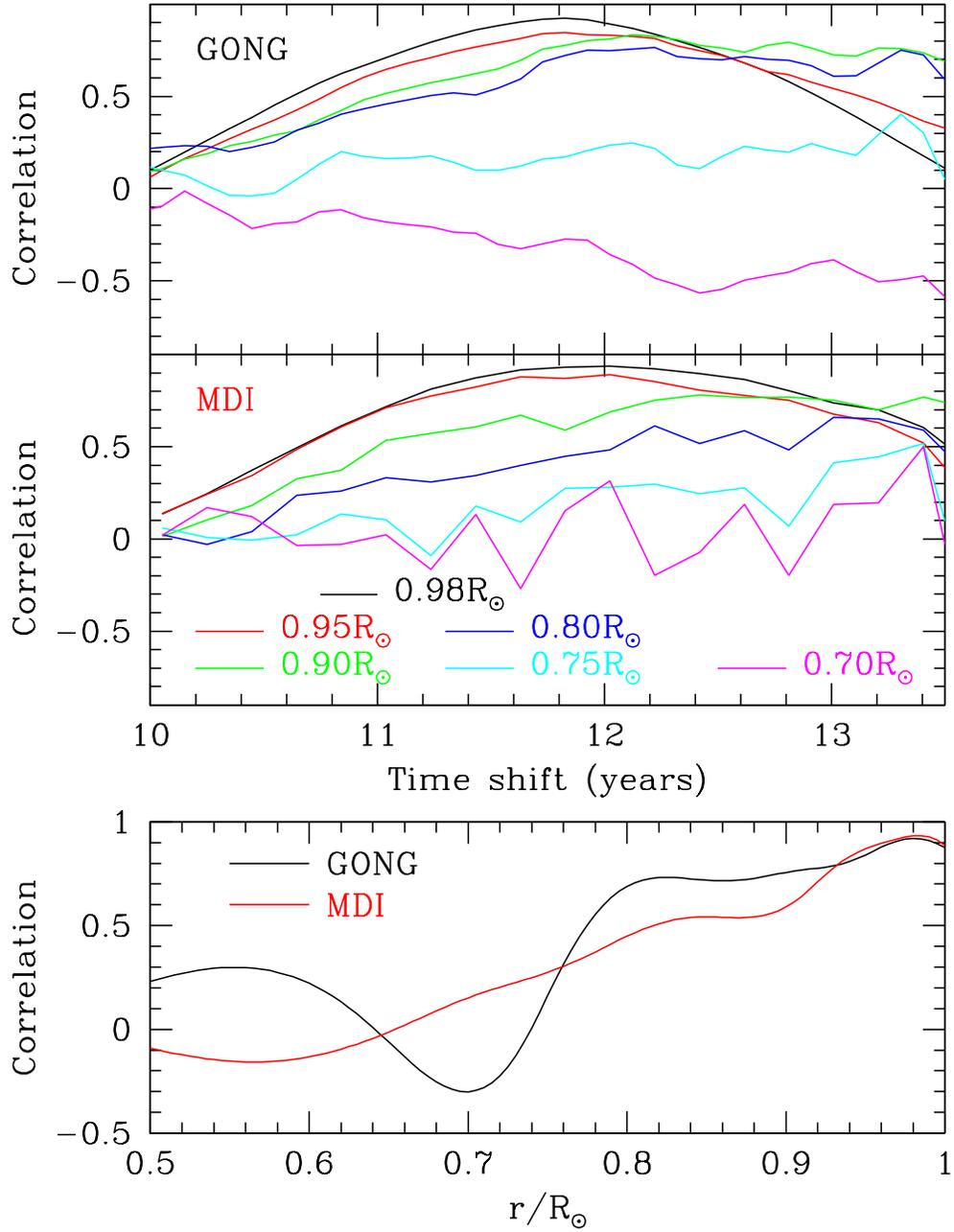}
\caption{The same as Figure~\ref{fig:cor}, but with summation restricted
to low latitude region ($\le 45^\circ$).}
\label{fig:corlow}
\end{figure}

The correlation coefficients obtained when we restrict the summation in Eq.~(\ref{eq:cor}) to the low-latitude regions
is shown in Figure~\ref{fig:corlow}.
It can be seen that for the GONG data these results are not very different
from those shown in Figure~\ref{fig:cor} that were obtained for the full latitude range.
The only difference is that the correlation coefficient for the low-latitude regions
is generally higher and the drop near the base of the
convection zone is sharper. 
The value of $T$ for which the  correlation coefficient is highest
does not change much with depth. The peak occurs at $T=11.7$ years.
Since the low-latitude pattern of zonal flows
is more well defined than the high-latitude pattern, it is perhaps more
 meaningful to restrict the analysis to the low-latitude regions.
For the  low-latitude region  MDI data result in smoother curves that are very similar to
those for GONG. The peak in MDI data also occurs at the same value of $T$
while considering only low latitude region.
On the other hand, if the summation is restricted to high
latitudes, then the variation of the correlation coefficient
with $T$ is similar to that obtained using both low- and high-latitude regions.
This is probably because zonal flow velocities are generally higher in the high-latitude
regions, thereby giving larger contribution to the summation.
Thus most of the difference between GONG and MDI data appears to be for
high latitudes.
Since the zonal-flow pattern is better defined in the low latitude regions
near the surface we use the estimate of period in this region to represent the 
length of solar cycle 23 defined by solar dynamics. This value is around 11.7 years.

Although it is difficult to define the period of minimum activity using
zonal-flow pattern, once the length of solar cycle 23 is estimated we
can use the known epoch of the minimum (1996.4) to calculate the epoch
of recent minimum. This turns out to be 2008.1, which is consistent with
time at which the high latitude band of faster than average rotation ends.
This is when the minimum in sunspot numbers is used to define the minimum of
solar cycle 23. If instead the 10.7 cm solar radio flux is used to define
the minimum, the estimate for current minimum will shift to 2008.5. The
minimum in sunspot number or 10.7 cm radio flux for solar cycle 24 occurred a few months later.
Thus the solar dynamics seems to suggest a slightly smaller period 
for solar cycle 23  compared to some other activity indices. 
Nevertheless, the difference is within the variations between different
activity indices.
Our estimate of length of the last solar cycle is consistent with that
obtained by Howe et al.~(2009) also from the study of zonal flows.

Figure~\ref{fig:difcut} which shows the difference in rotation velocity
at two different epochs, can actually be used
to check our estimate of the length of cycle 23 using zonal flows.
It can be seen that in the low-latitude near-surface layers the
differences between 1996.4 and 2008.1 are smaller than those
between 1996.4 and 2008.9.
The epoch of 2008.9 is estimated to be the minimum as measured by solar
activity indices (Hathaway 2010).
Of course, the differences are nonzero in both cases as there are significant 
differences in the zonal-flow patterns during the two epochs.

\section{Conclusions}
\label{sec:disc}

The recent cycle 24 minimum in solar activity is known to be unusual in 
terms of extended duration and abnormally low activity levels. 
We have used helioseismic data from the GONG and MDI projects to study 
whether there were differences in solar dynamics between the
minimum of solar cycle 23 and that of cycle 24.

We find that there are substantial differences between the rotation
rates during the two minima. There are differences throughout
the convection zone, including the tachocline and 
the outer shear layer.
Even though the rotational velocities are different in the
outer shear layer, the extent of the shear layer and the increase in the
rotation rate across the shear layer do not appear to change with time.

The differences between the two minima are more obvious if one looks at the
zonal-flow pattern. In addition to the slowing down of the mid-latitude
prograde band that was reported by Howe et al.~(2009), we find that the
low-latitude prograde band that bifurcated in 2005  merged well before the
cycle 24 minimum. This is in contrast to this band's behavior 
during the cycle 23 minimum where the merging happened just before the
minimum.
Since our inversion results are sensitive only to the north-south symmetric
component of the rotation rate, it is possible that the splitting of the bands is 
actually a near-equator flow in one hemisphere only.
This band was stronger before the cycle 24 minimum than
it was in the cycle 23 case. The high-latitude prograde band was also different.
During the Cycle 23 minimum it had ended around 1996.5, while in the
cycle 24 case, the band ended around 2008.0 although solar activity levels
continued to be low for about a year after that. Furthermore, the
mid-latitude bands evolved slower before the cycle 24 minimum and the
low latitude band appears to be extended over longer time after merger during
cycle 24 probably reflecting the extended minimum in activity.

If we assume that the zonal flows show a cyclical behavior in the same
way as the magnetic indices do, we can determine the period of the cycle 
by correlating the zonal-flow pattern with itself after a shift in time.
Using this we estimate the
length of solar cycle 23 to be about 11.7 years. This is shorter than the
length defined as the time elapsed between the minimum in solar activity  indices.
For cycle 23 the period derived from sunspot numbers is about 12.6 years
(see Hathaway 2010).

We conclude by reiterating that solar dynamics, in particular the
solar rotation rate and zonal flows were different for the cycle 24
minimum compared with the cycle 23 minimum. 
We find that solar zonal flows returned to
their mean-minimum state before solar magnetic indices did during the end of
solar cycle 23.

\acknowledgements

This work  utilizes data obtained by the Global Oscillation
Network Group (GONG) project, managed by the National Solar Observatory,
which is
operated by AURA, Inc. under a cooperative agreement with the
National Science Foundation. The data were acquired by instruments
operated by the Big Bear Solar Observatory, High Altitude Observatory,
Learmonth Solar Observatory, Udaipur Solar Observatory, Instituto de
Astrofisico de Canarias, and Cerro Tololo Inter-American Observatory.
This work also utilizes data from the Solar Oscillations
Investigation/ Michelson Doppler Imager (SOI/MDI) on the Solar
and Heliospheric Observatory (SOHO).  SOHO is a project of
international cooperation between ESA and NASA.
SB acknowledges support from NSF grant ATM 0348837
and NASA grant NXX10AE60G.

\clearpage

\clearpage

\clearpage

\clearpage

\clearpage

\clearpage

\clearpage

\clearpage

\clearpage

\end{document}